\documentclass[twocolumn, switch]{article} 

\usepackage{preprint}

\usepackage{amsmath, amsthm, amssymb, amsfonts}

\usepackage[numbers,square]{natbib}

\usepackage[utf8]{inputenc}	
\usepackage[T1]{fontenc}	
\usepackage{xcolor}		
\usepackage[colorlinks = true,
            linkcolor = purple,
            urlcolor  = blue,
            citecolor = cyan,
            anchorcolor = black]{hyperref}	
\usepackage{booktabs} 		
\usepackage{nicefrac}		
\usepackage{microtype}		
\usepackage{lineno}		
\usepackage{float}			

\usepackage{lipsum}		

\usepackage{newfloat}
\DeclareFloatingEnvironment[name={Supplementary Figure}]{suppfigure}
\usepackage{sidecap}
\sidecaptionvpos{figure}{c}

\usepackage{titlesec}
\titlespacing\section{0pt}{12pt plus 3pt minus 3pt}{1pt plus 1pt minus 1pt}
\titlespacing\subsection{0pt}{10pt plus 3pt minus 3pt}{1pt plus 1pt minus 1pt}
\titlespacing\subsubsection{0pt}{8pt plus 3pt minus 3pt}{1pt plus 1pt minus 1pt}

\usepackage[utf8]{inputenc}
\usepackage{amsmath}
\usepackage{graphicx}
\usepackage{wrapfig}
\usepackage{float}
\usepackage{arydshln}
\usepackage{todonotes}
\usepackage[capitalize,noabbrev]{cleveref}
\usepackage[labelformat = parens,justification=centering]{subfig}
\usepackage{xspace}

\newcommand{\spmv}{\texttt{SpMV}\xspace}
\renewcommand{\ell}{\texttt{ELL}\xspace}

\title{Evaluating the 
Performance of NVIDIA's A100 Ampere GPU\\ for Sparse Linear 
Algebra Computations}

\usepackage{xwatermark}

\author{Yuhsiang Mike Tsai \\
 Karlsruhe Institute of Technology, \\
 Germany\\
 \texttt{yu-hsian.tsai@kit.edu}
\And
Terry Cojean \\
 Karlsruhe Institute of Technology, \\
 Germany\\
 \texttt{terry.cojean@kit.edu}
\And
Hartwig Anzt \\
 Karlsruhe Institute of Technology, \\
 Germany\\
 University of Tennessee, Knoxville, \\
 USA\\
 \texttt{hartwig.anzt@kit.edu}
}

\begin{document}

\twocolumn[ 
  \begin{@twocolumnfalse} 
  
\maketitle

\begin{abstract}
    GPU accelerators have become an important backbone for scientific high 
    performance computing, and the performance advances obtained from adopting new 
    GPU hardware are significant. In this paper we take a first look at NVIDIA's 
    newest server line GPU, the A100 architecture part of the Ampere generation. 
    Specifically, we assess its performance for sparse linear algebra operations 
    that form the backbone of many scientific applications and assess the 
    performance improvements over its predecessor.
\end{abstract}
\keywords{Sparse Linear Algebra \and Sparse Matrix Vector Product \and NVIDIA A100 GPU} 
\vspace{0.35cm}

  \end{@twocolumnfalse} 
] 


\section{Introduction}
Over the last decade, Graphic Processing Units (GPUs) have seen an increasing 
adoption in high performance computing platforms, and in the June 2020 TOP500 
list, more than half of the fastest 10 systems feature GPU 
accelerators~\cite{top500}. At the same time, the June 2020 edition of the 
TOP500 is the first edition listing a system equipped with NVIDIA's new A100 
GPU, the HPC line 
GPU of the Ampere generation. As the scientific high performance computing 
community anticipates this GPU to be the new flagship architecture in 
NVIDIA's hardware portfolio, we take a look at the performance we achieve on 
the A100 for sparse linear algebra operations.

Specifically, we first 
benchmark in \Cref{sec:bandwidth} the bandwidth of the A100 GPU for 
memory-bound vector operations and compare against NVIDIA's A100 predecessor, 
the V100 GPU. In \Cref{sec:spmv}, we review the sparse matrix vector product 
(\spmv), a central kernel for sparse linear algebra, and outline the processing 
strategy used in some popular kernel realizations. In 
\Cref{sec:spmvexperiments}, we 
evaluate the performance of \spmv kernels on the A100 GPU for more than 2,800 
matrices available in the Suite Sparse Matrix Collection~\cite{suitesparse}. 
The \spmv kernels we consider in this performance evaluation are taken from 
NVIDIA's latest release of the cuSPARSE library and the Ginkgo linear algebra 
library~\cite{ginkgo}. 
In \Cref{sec:krylov}, we compare the performance of the A100 against its 
predecessor for complete Krylov solver iterations that are popular methods for 
iterative sparse linear system solves.
In \Cref{sec:conclusion}, we summarize the performance assessment results and
draw some preliminary conclusions on what performance we may achieve for sparse
linear algebra on the A100 GPU.

We emphasize that with this paper, we do not intend to provide another 
technical specification of NVIDIA's A100 GPU, but instead focus on the 
reporting of performance we observe on this architecture for sparse linear 
algebra operations. 
Still, for convenience, we append a table from NVIDIA's whitepaper on the 
NVIDIA A100 Tensor Core GPU Architecture~\cite{a100} that lists some key 
characteristics and compares against the predecessor GPU architectures. 
For further information on the A100 GPU, we refer to the 
whitepaper~\cite{a100}, and encourage the reader to digest the performance 
results we present side-by-side with these technical details.

\section{Memory Bandwidth Assessment}
\label{sec:bandwidth}
The performance of sparse linear algebra operations on modern hardware
architectures is usually limited by the data access rather than compute power.
In consequence, for sparse linear algebra, the performance-critical hardware
characteristics are the memory bandwidth and the access latency. For main 
memory access, both metrics are typically somewhat intertwined, in particular 
on processors operating in streaming mode like GPUs. 
In this section, we assess the memory access performance by 
means of the Babel-STREAM benchmark~\cite{babelstream}.

\begin{figure*}[!h]
    \centering
    \subfloat[V100\label{fig:v100_bandwidth}
    ]{\includegraphics[width=.45\textwidth]{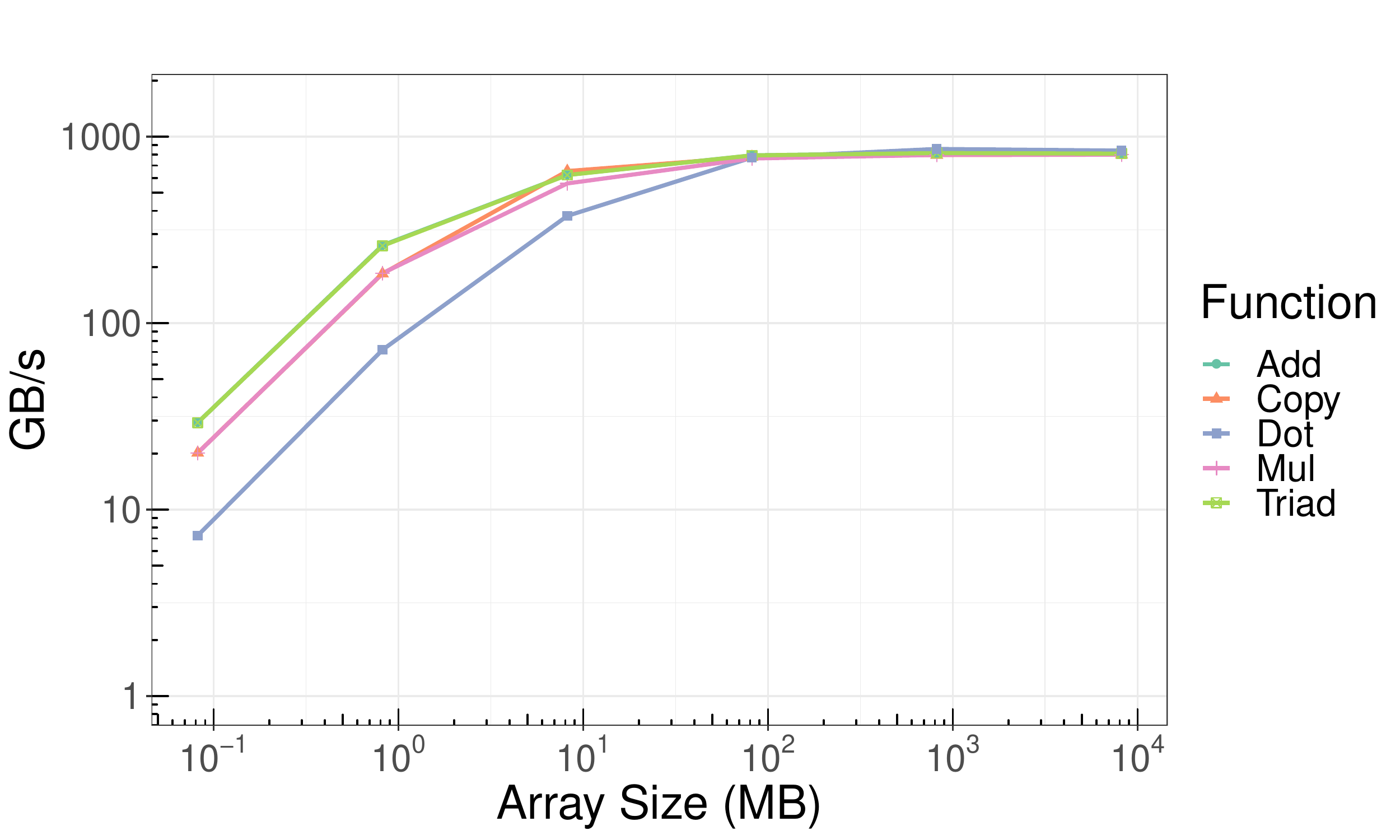}
    }
    \subfloat[A100\label{fig:a100_bandwidth}
    ]{\includegraphics[width=.45\textwidth]{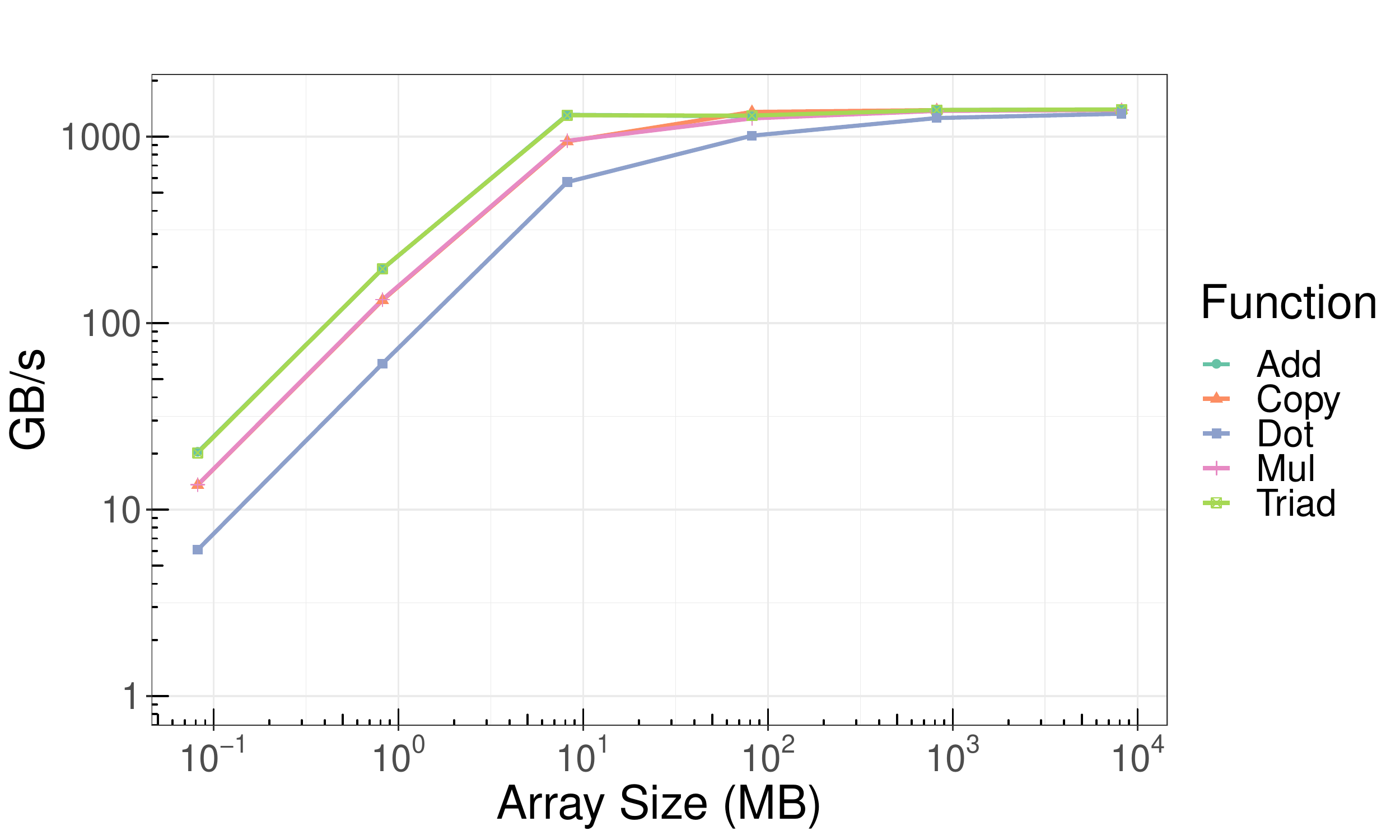}
    }
    \caption{Performance of the Babel-STREAM benchmark on both V100 and A100
      NVIDIA GPUs.}
    \label{fig:a100_v100_bandwidth}
\end{figure*}

We show the Babel-STREAM benchmark results for
both an NVIDIA V100 GPU \Cref{fig:v100_bandwidth} and an NVIDIA A100 GPU
\Cref{fig:a100_bandwidth}. The figures reflect a significant bandwidth 
improvement
for all operations on the A100 compared to the V100. For an array of size 8.2~GB,
the V100 reaches for all operations a performance between 800 and 840 GB/s
whereas the A100 reaches a bandwidth between 1.33 and 1.4 TB/s.

\begin{figure*}[!h]
    \centering
    \subfloat[Evolution of the A100/V100 ratio for different array sizes.\label{fig:a100_v100_ratio}
    ]{\includegraphics[width=.45\textwidth]{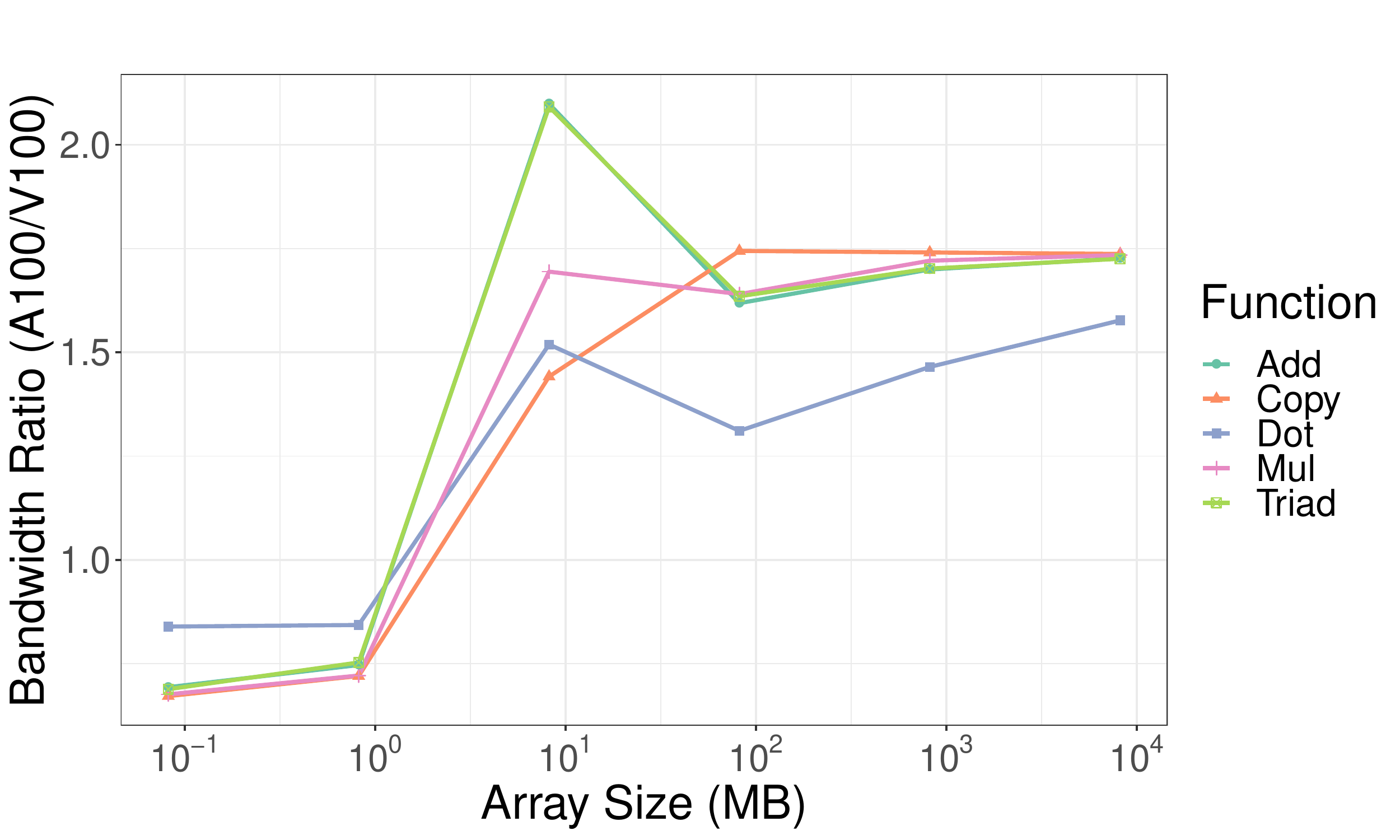}
    }
    \subfloat[A100/V100 ratio for an array of size 8.2~GB.\label{fig:a100_v100_max_ratio}
    ]{\includegraphics[width=.45\textwidth]{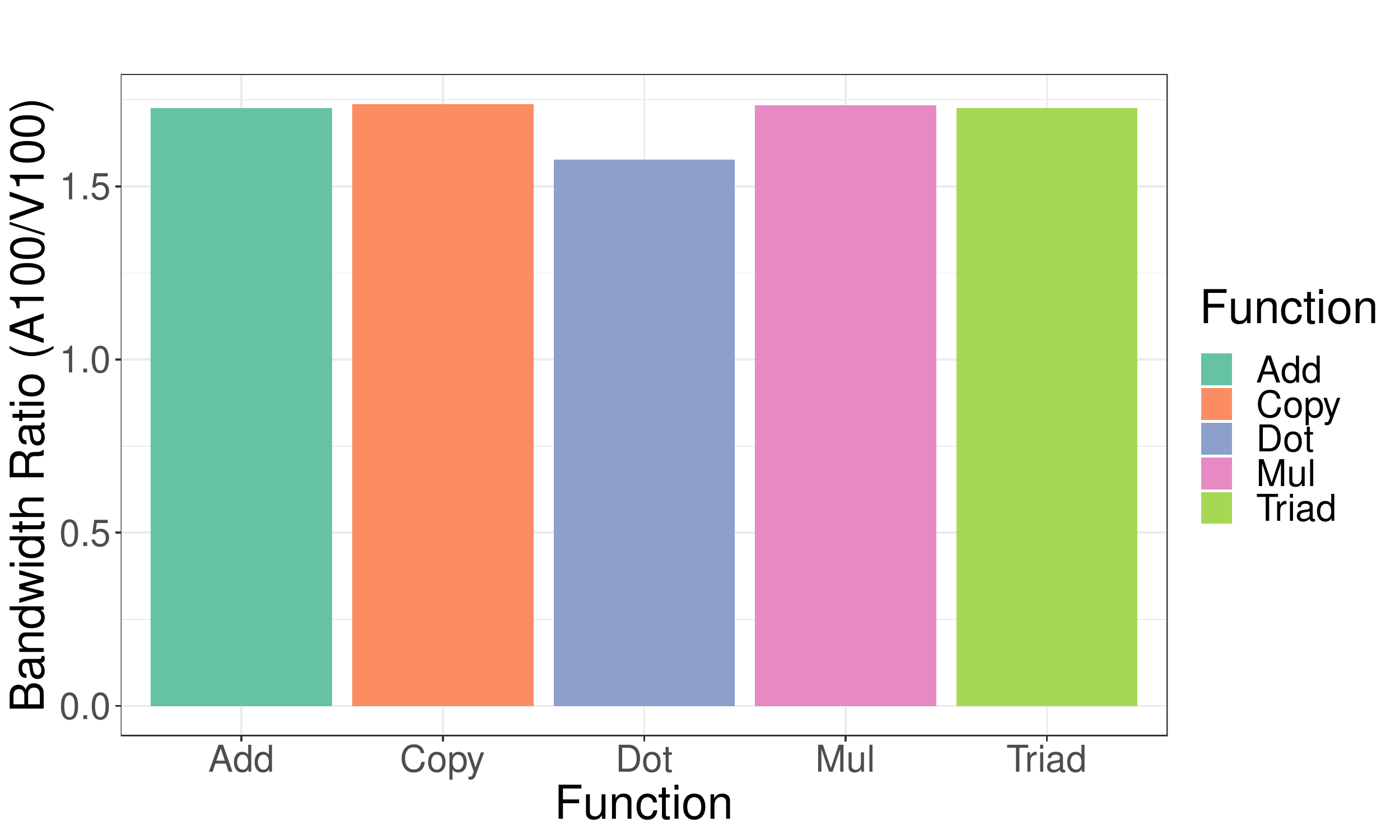}
    }
    \caption{Ratio of the performance improvement of A100 compared to V100
      NVIDIA GPUs for the Babel-STREAM benchmark.}
    \label{fig:a100_v100_bandwidth_ratio}
\end{figure*}

In \Cref{fig:a100_v100_bandwidth_ratio}, we present the data as a ratio between 
the
A100 and V100 bandwidth performance for all operations. In
\Cref{fig:a100_v100_ratio}, we show the performance ratio for increasing array 
sizes and observe the A100 providing a lower bandwidth than the V100 GPU for 
arrays of small size. For large array sizes, the A100 has a significantly 
higher bandwidth, converging towards a speedup factor of 1.7 for most memory 
access benchmarks, see \Cref{fig:a100_v100_max_ratio}.
\begin{figure*}
\centering
	\includegraphics[width=1.0\linewidth]{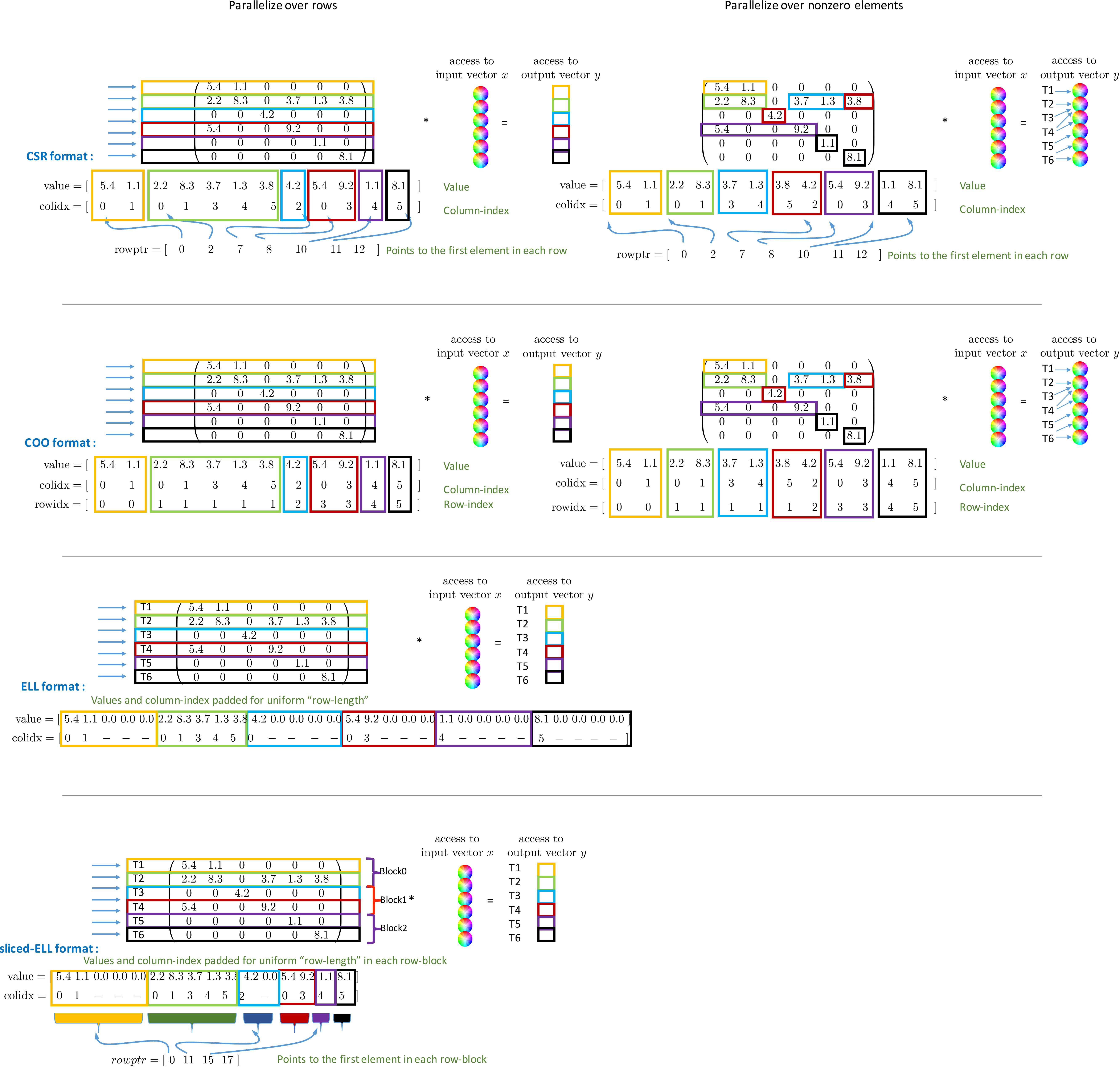}
    \caption{Overview over sparse matrix formats and \spmv kernel design.}
    \label{fig:spmvoverview}
\end{figure*}

\begin{figure*}[!h]
    \centering
    \subfloat[\spmv performance profile on A100.\spmv\label{fig:a100spmv}
    ]{\includegraphics[width=.45\textwidth]{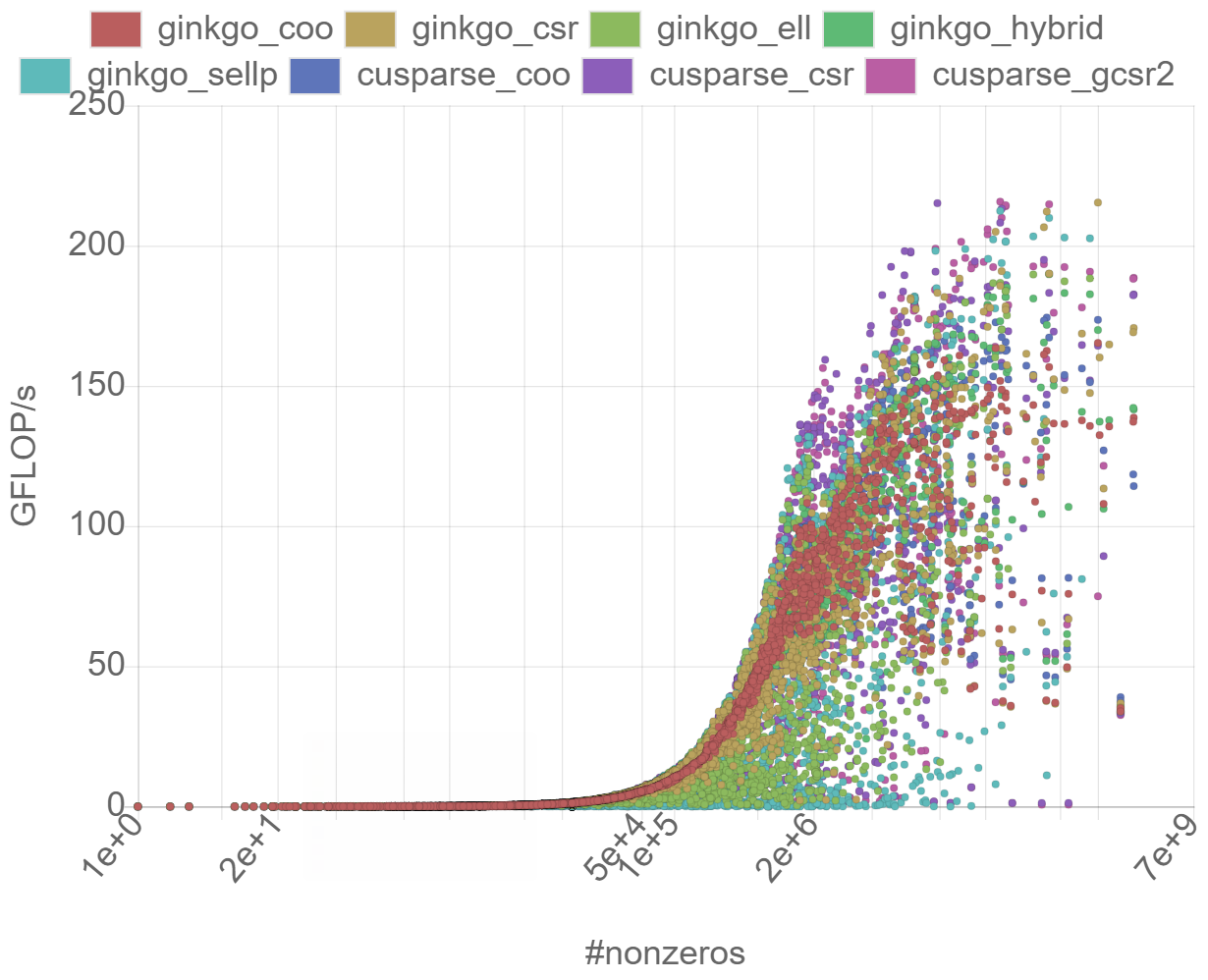}
    }
    \subfloat[Performance profile on A100.\label{fig:a100profile}
    ]{\includegraphics[width=.45\textwidth]{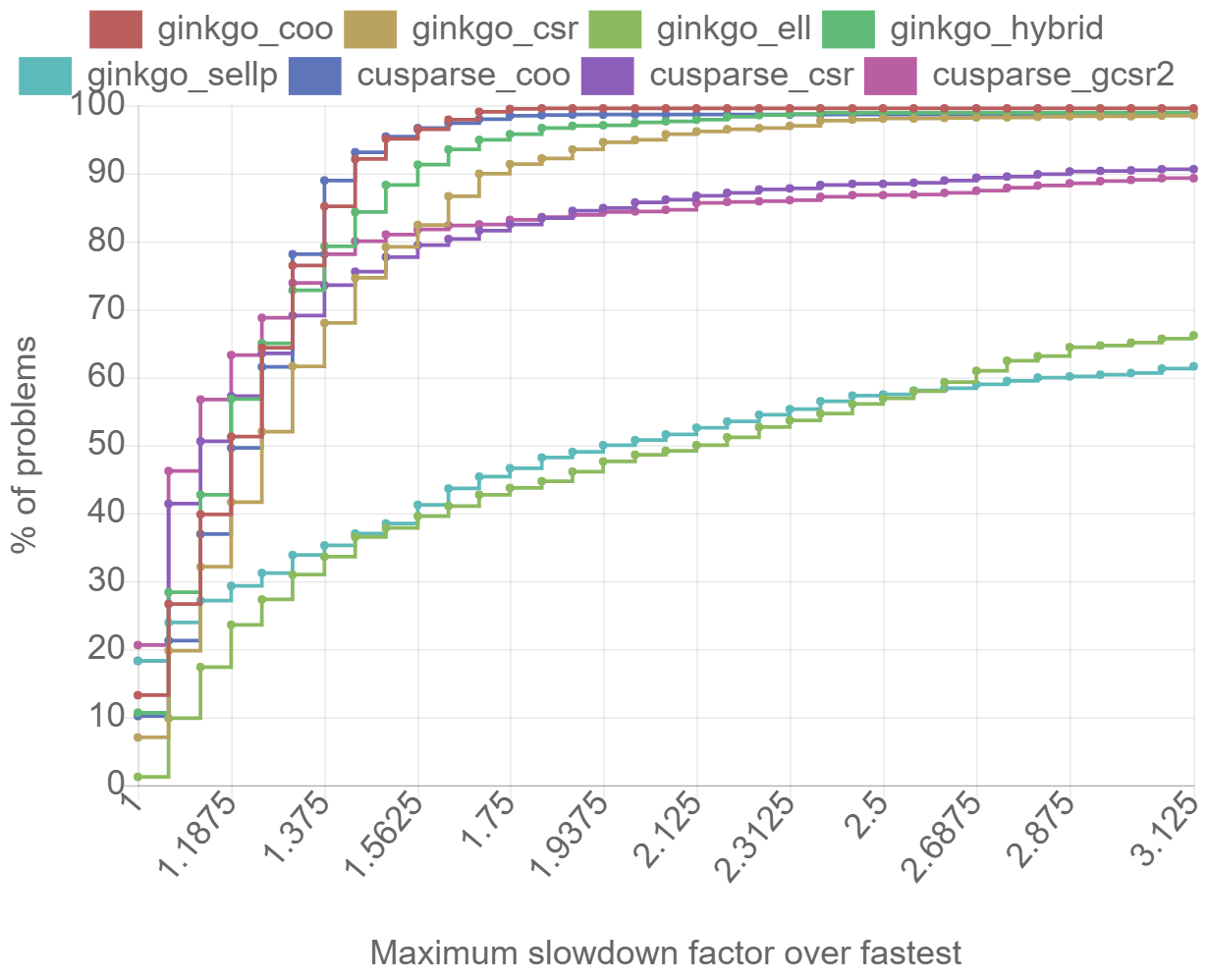}
    }
    \caption{Left: \spmv kernel performance on the A100 GPU considering 2,800 
    test matrices from the Suite Sparse Matrix Collection. Right: 
    Corresponding Performance profile for all \spmv kernels considered.}
    \label{fig:spmvperf}
\end{figure*}

\section{Sparse Matrix Vector Product}
\label{sec:spmv}
The sparse matrix-vector product (\spmv) is a heavily-used operation in many 
scientific applications. The spectrum ranges from the power 
iteration~\cite{poweriteration}, an iterative algorithm for finding eigenpairs 
in Google's Page Rank algorithm~\cite{Langville:2012:GPB:2331354}, to iterative 
linear system solvers like Krylov solvers that form the backbone of many finite 
element simulations.
Given this importance, we put particular focus on the performance of the \spmv
operation on NVIDIA's A100 GPU. However, the performance not only depends on the
hardware and the characteristics of the sparse matrix but also the specific
\spmv format and processing strategy. 
Generally, all \spmv kernels aim at reducing the memory access
cost (and computational cost) by storing only the nonzero matrix
values~\cite{barrettemplates}. Some formats additionally store a moderate amount
of zero elements to enable faster processing when computing matrix-vector
products~\cite{ellpack}. But independent of the specific strategy, since \spmv
kernels store only a subset of the elements, they share the need to accompany
these values with information that allows to deduce their location in the
original matrix~\cite{spmvtopc}. We here recall in \Cref{fig:spmvoverview} some
widespread sparse matrix storage formats and kernel parallelization techniques.

A straightforward idea is to accompany the nonzero elements with the 
respective row and column indexes.
This storage format, known as coordinate (COO~\cite{barrettemplates}) format,
allows determining the original position of any element
in the matrix without processing other entries, see first row in 
\Cref{fig:spmvoverview}. A standard parallelization of 
this approach assigns the matrix rows to the distinct processing elements 
(cores). However, if few rows contain a significant portion of the overall 
nonzeros, this can result in a significant imbalance of the kernel and poor 
performance. A workaround is to distribute the nonzeros across the parallel 
resources, see the right-hand side in \Cref{fig:spmvoverview}. However, as this 
can 
result in 
several processing elements 
contributing partial sums to the same vector output entry, sophisticated 
techniques for lightweight synchronization are needed to avoid write 
conflicts~\cite{coopaper}.

Starting from the COO format, further reduction of the storage cost is possible 
if the elements are sorted row-wise, and with increasing column-order in every 
row.
(The latter assumption is technically not required,
but it usually results in better performance.)
Then, this Compressed Sparse Row (CSR~\cite{barrettemplates}) format
can replace the array containing the row indexes with a pointer to the
beginning of the distinct rows, see the second row in \Cref{fig:spmvoverview}. 
While 
this generally reduces the data volume,
the CSR format requires extra processing to determine the row location of
a certain element. For a standard parallelization over the matrix rows, the row
location is implicitly given, and no additional information is needed. However,
similar to the COO format, better load balancing is available if parallelizing
across nonzero elements. This requires the matrix row information and
sophisticated atomic writes to the output vector~\cite{csri}.

A strategy that reduces the row indexing information even further is to pad 
all rows to the same number of nonzero elements and accompany the values only 
with the column indexes. In this ELL format~\cite{ellpack} (third row in 
\Cref{fig:spmvoverview}), the row index of an 
element can be deduced from its location in the array storing the values in
consecutive order and the information of how many elements are stored in each
row. While this format is attractive for vector processors as it allows to
execute in SIMD fashion with coalesced memory access, its efficiency heavily
depends on the matrix characteristics: for well-balanced matrices, it optimizes
memory access cost and operation count; but if one or a few rows contain much
more nonzero elements than the others, ELL introduces a significant padding
overhead and quickly becomes inefficient~\cite{spmvtopc}.

An attractive strategy to reduce the padding overhead the ELL format introduces 
for unbalanced matrices is to decompose the original matrix into blocks 
containing multiple rows, and to store the distinct blocks in ELL format. Rows 
of the same block contain the same number of nonzero elements, rows in distinct 
blocks can differ in the number of nonzero elements. 
In this Sliced ELL format (SELL~\cite{kreutzer}, see the fourth row in 
\Cref{fig:spmvoverview}), the row pointer can not 
completely be omitted 
like in the ELL case, but a row pointer to the beginning of every block is 
needed. In this sense, the SELL format is a trade-off between the ELL format on 
the one side and the CSR format on the other side. Indeed, choosing a block 
size of 1 results in the CSR format, choosing a block size of the matrix size 
results in the ELL format. In practice, the block size is adjusted to the 
matrix properties and the characteristics of the parallel hardware, i.e. 
SIMD-with~\cite{sellpreport}.

Another strategy to balance between the effectiveness of the ELL \spmv kernel
and the more general CSR/COO \spmv kernels is to combine the formats in a
``hybrid'' \spmv kernel~\cite{spmvtopc}. The concept behind is to store the
balanced part of the matrix in ELL format and the unbalanced part in CSR or COO
format. The \spmv operation then invokes two kernels, one for the balanced part
and one for the unbalanced part.

For all these \spmv kernel strategies, there has been significant research 
efforts to optimize their performance on GPU architectures, see, 
e.g.,~\cite{spmvtopc,
7161529,DBLP:journals/topc/MerrillGG15,Merrill:2016:MPS:3014904.3014982,
DBLP:conf/ppopp/HongSNSS19,Brin1998}
and references therein.
We here focus on the \spmv kernels of the cuSPARSE and Ginkgo libraries that 
are well-known to belong to the most efficient implementations available.

\begin{figure*}[!h]
  \centering
  \subfloat[cuSPARSE CSR \spmv\label{fig:cusparse_csr}
  ]{\includegraphics[width=.4\textwidth]{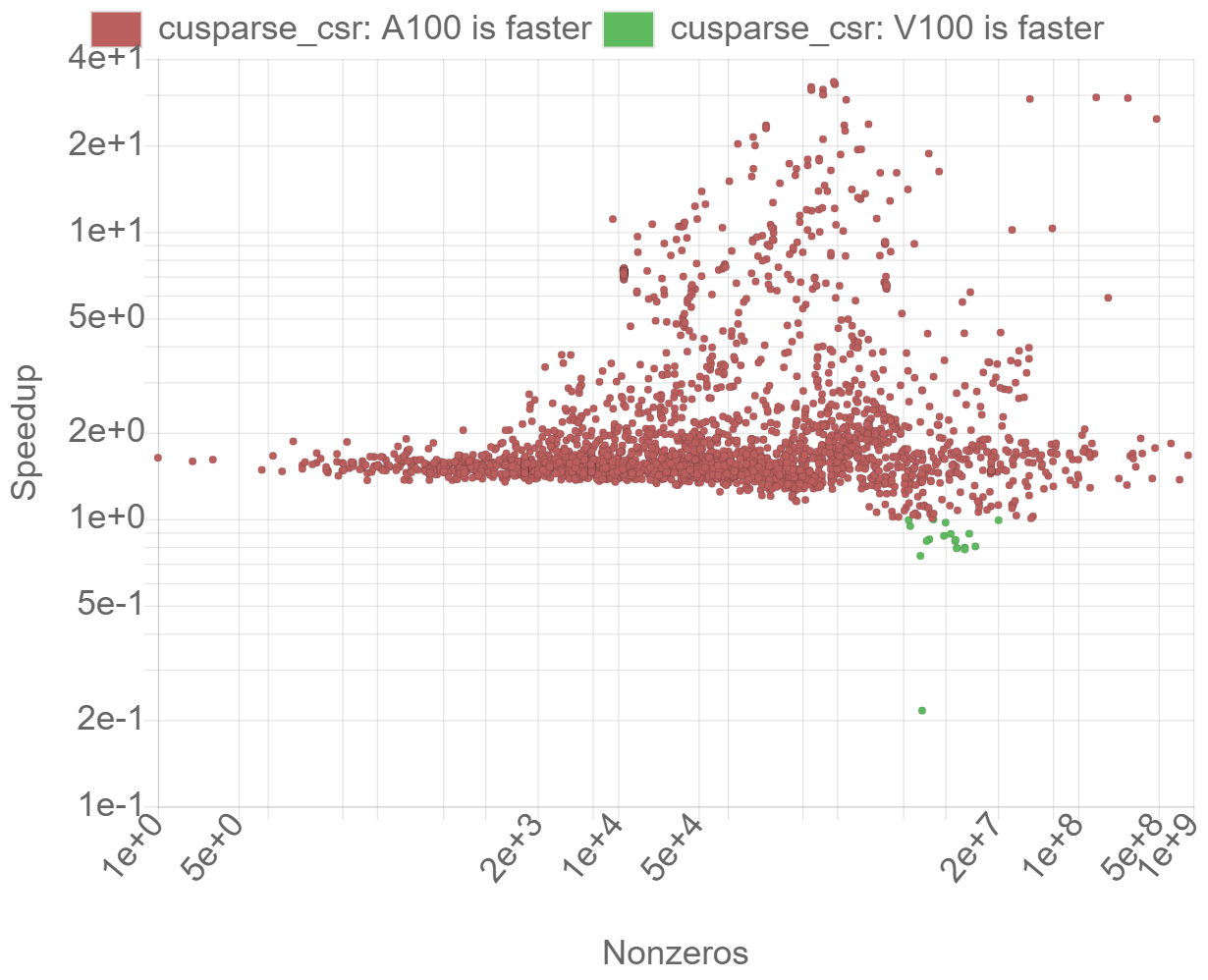}
  }
  \subfloat[Ginkgo CSR \spmv\label{fig:ginkgo_csr}
  ]{\includegraphics[width=.4\textwidth]{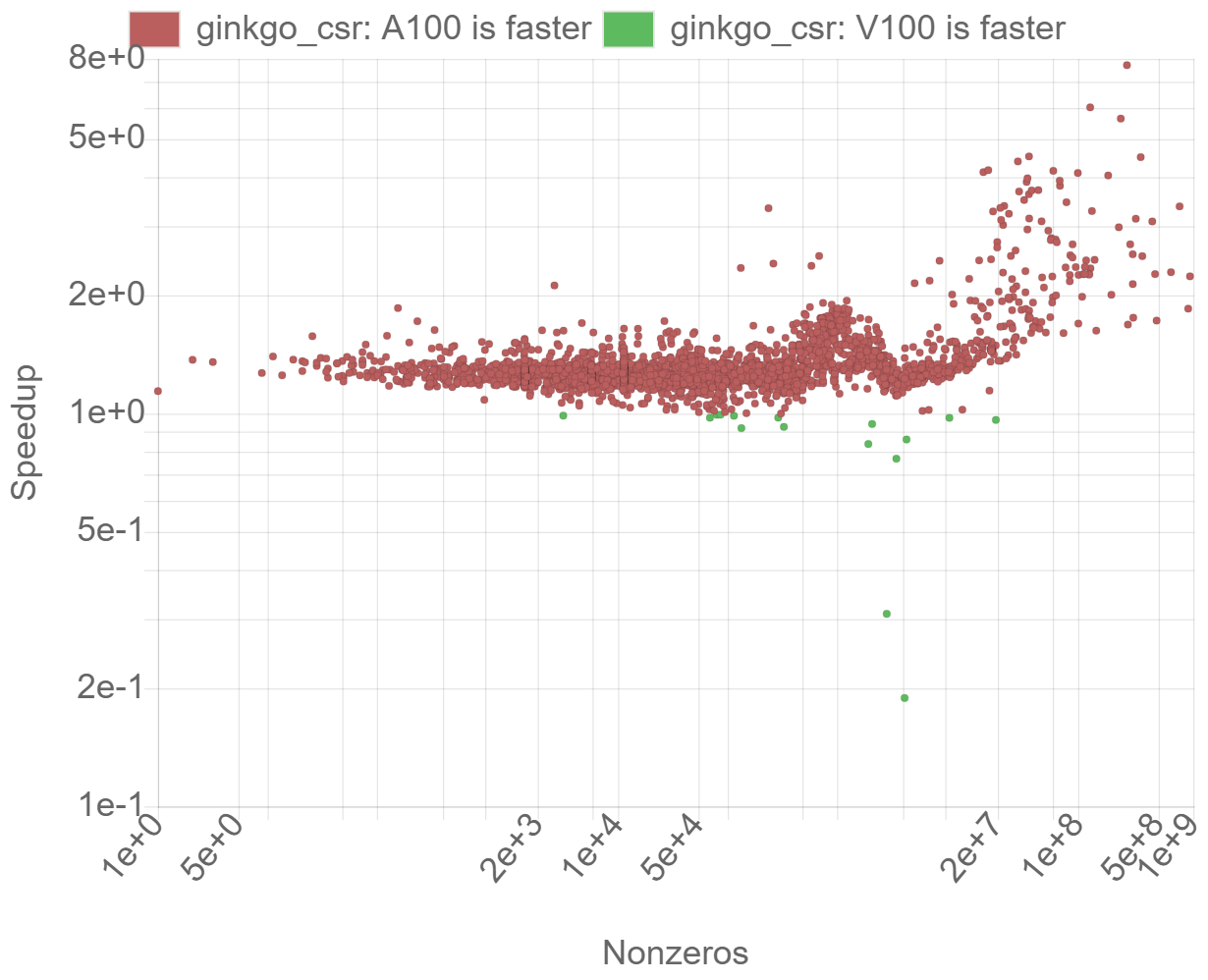}
  }\\
  \subfloat[cuSPARSE COO \spmv\label{fig:cusparse_coo}
  ]{\includegraphics[width=.4\textwidth]{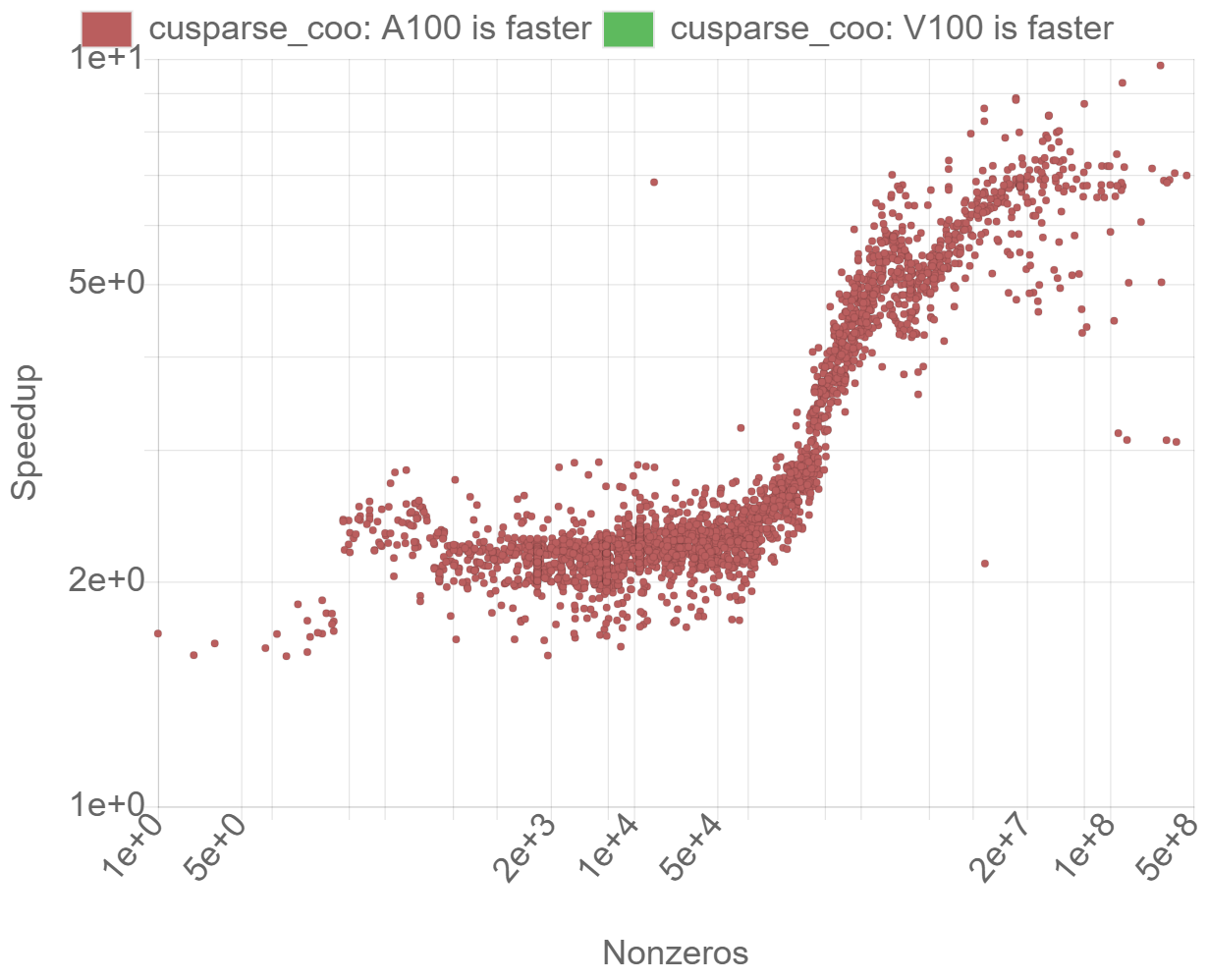}
  }
  \subfloat[Ginkgo COO \spmv\label{fig:ginkgo_coo}
  ]{\includegraphics[width=.4\textwidth]{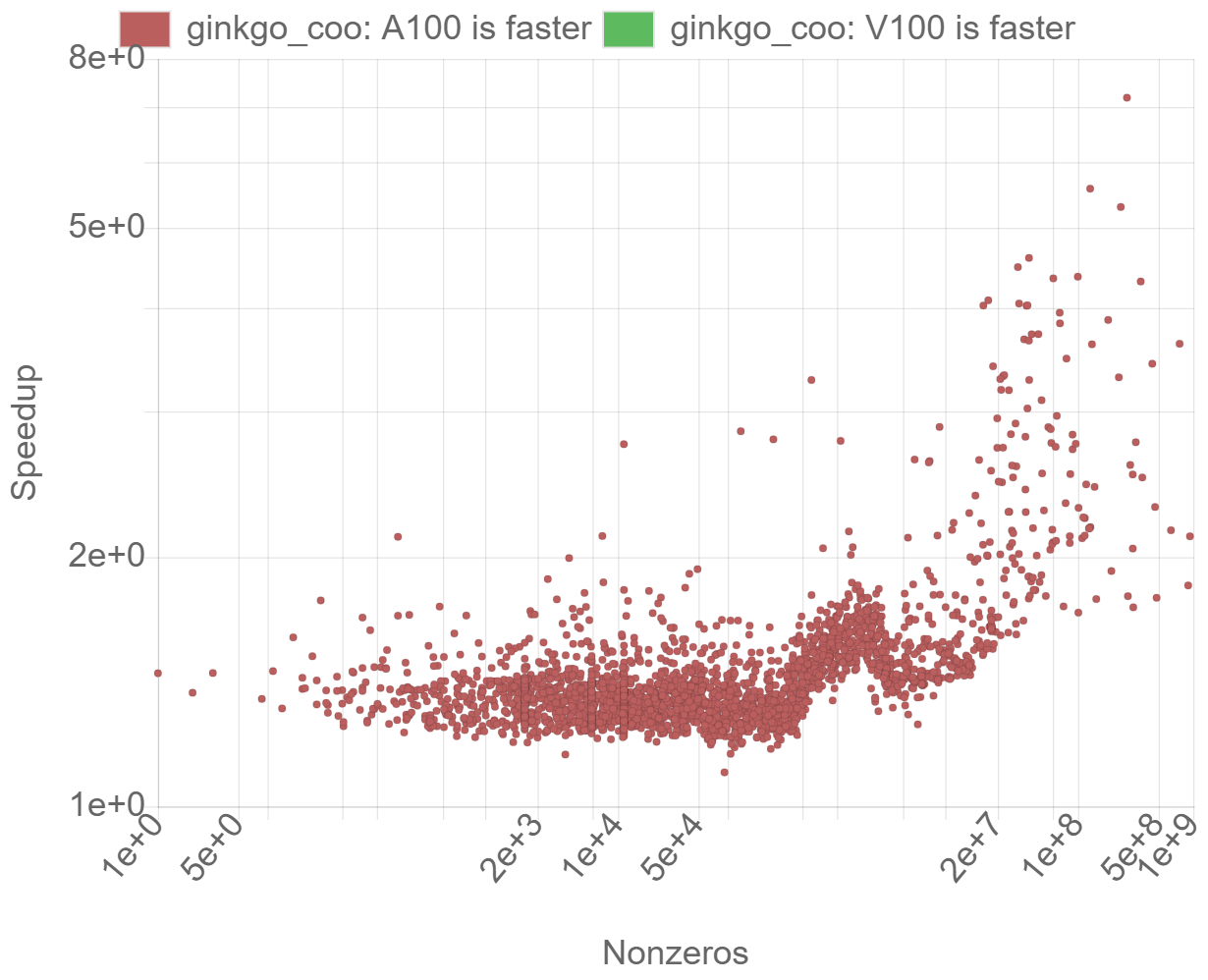}
  }\\
  \subfloat[Ginkgo ELL \spmv\label{fig:ginkgo_ell}
  ]{\includegraphics[width=.4\textwidth]{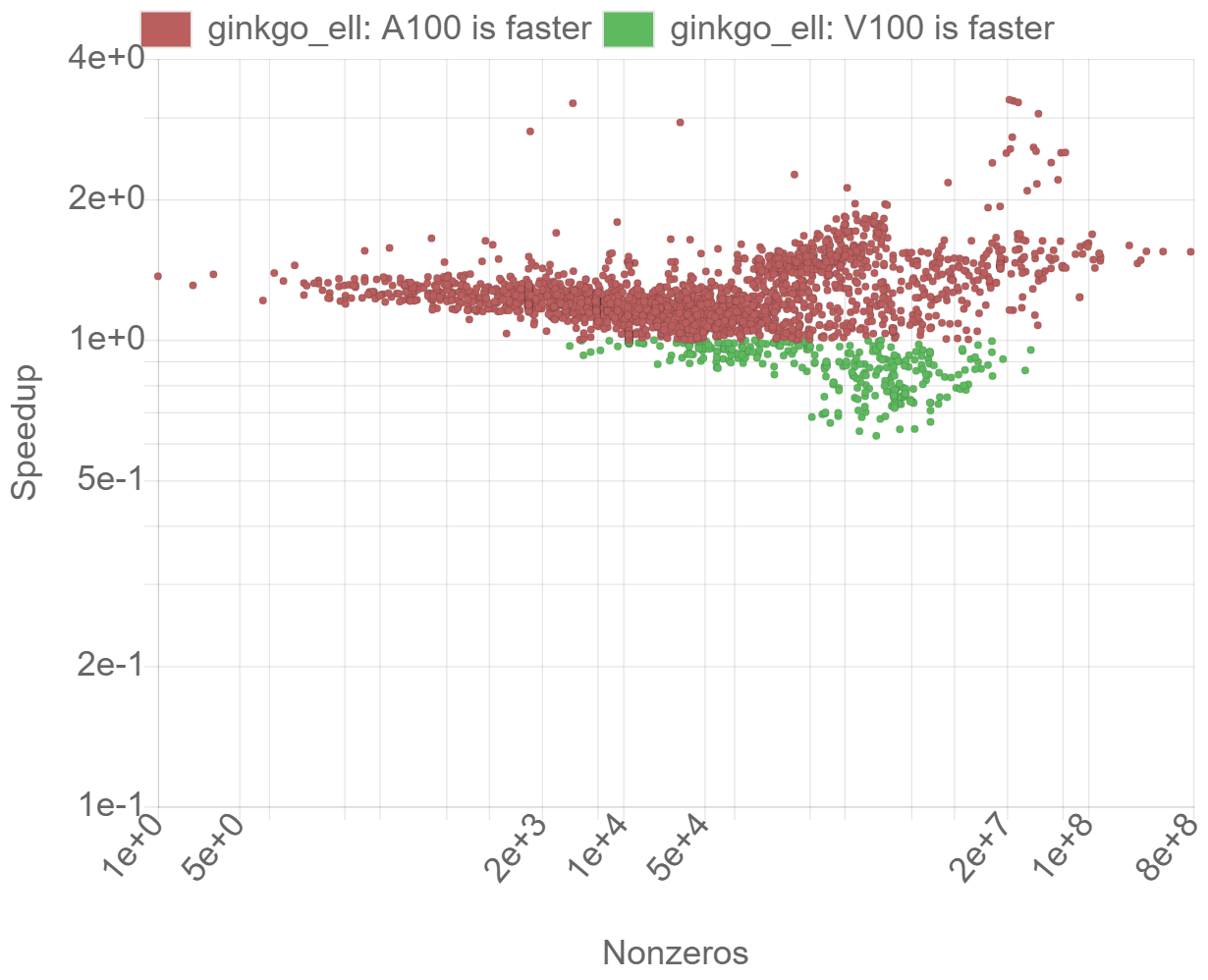}
  }
  \subfloat[Ginkgo hybrid \spmv\label{fig:ginkgo_hybrid}
  ]{\includegraphics[width=.4\textwidth]{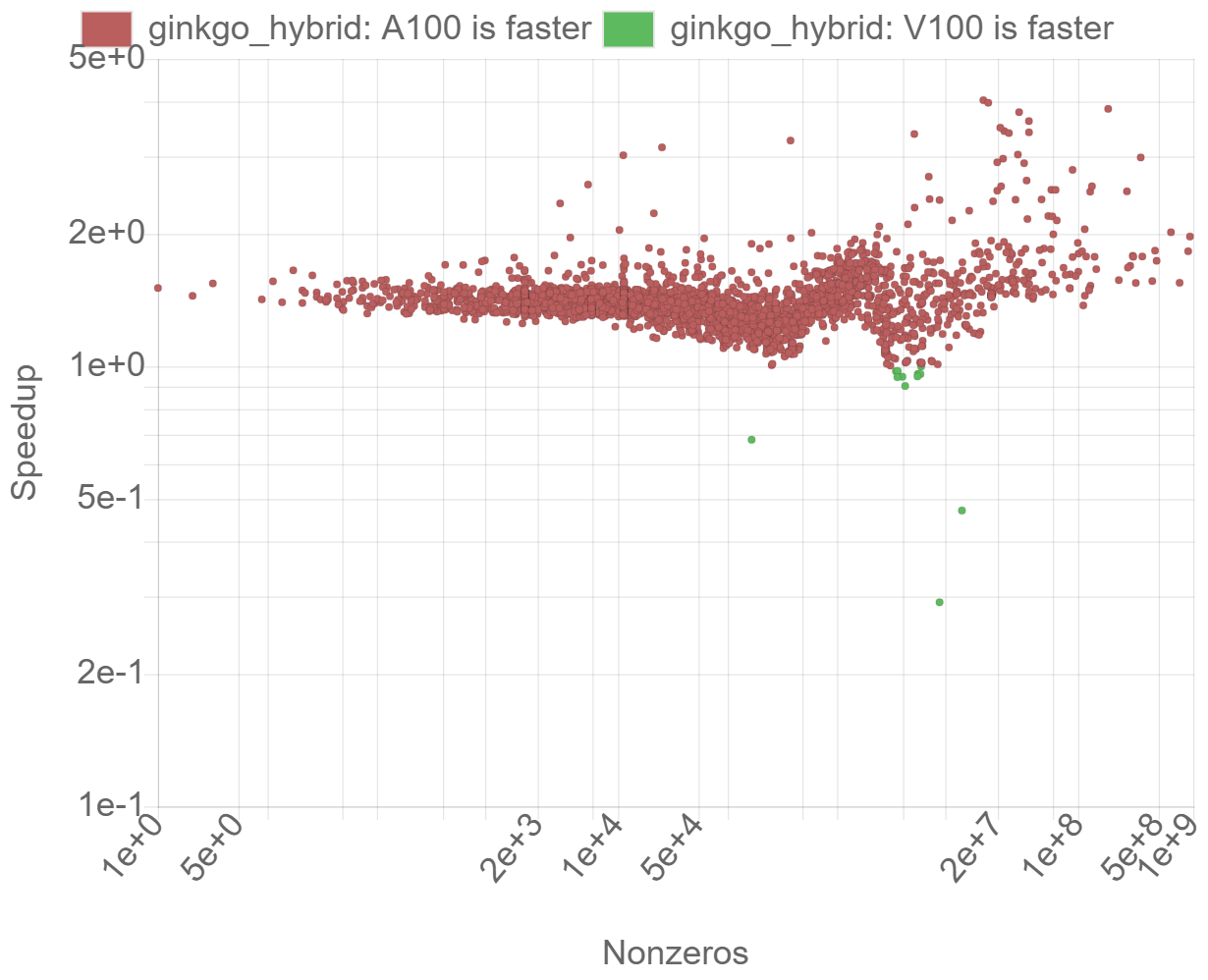}
  }
  \caption{Performance improvement assessment of the A100 GPU over the V100 
  GPU for \spmv kernels from NVIDIA's cuSPARSE library and Ginkgo.}
  \label{fig:v100a100spmv}
\end{figure*}

\begin{figure*}[!h]
  \centering
  \subfloat[Ginkgo Ell \spmv\label{fig:rcv_ell}
  ]{\includegraphics[width=.43\textwidth]{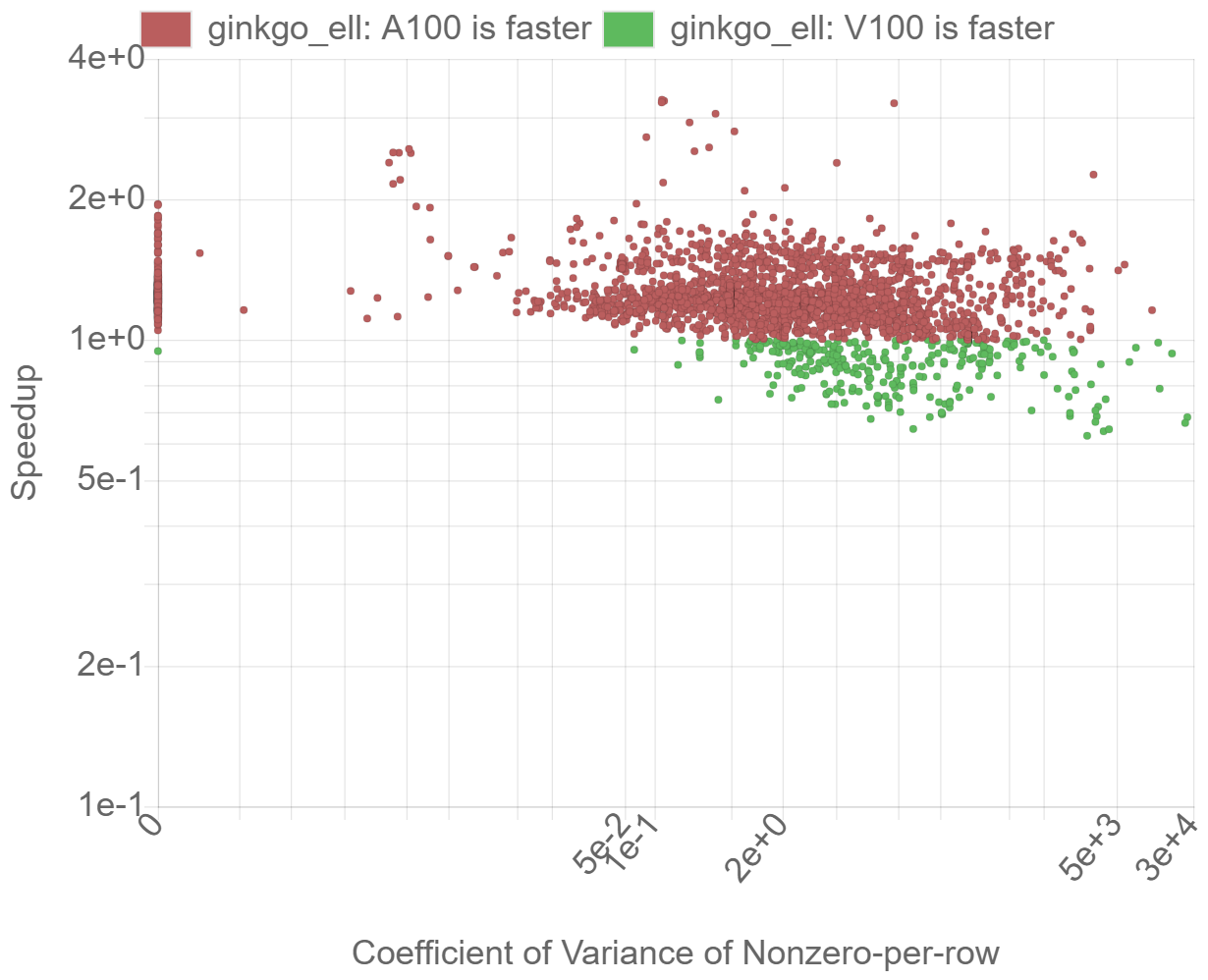}
  }
  \subfloat[Ginkgo Classical Csr \spmv\label{fig:rcv_csrc}
  ]{\includegraphics[width=.43\textwidth]{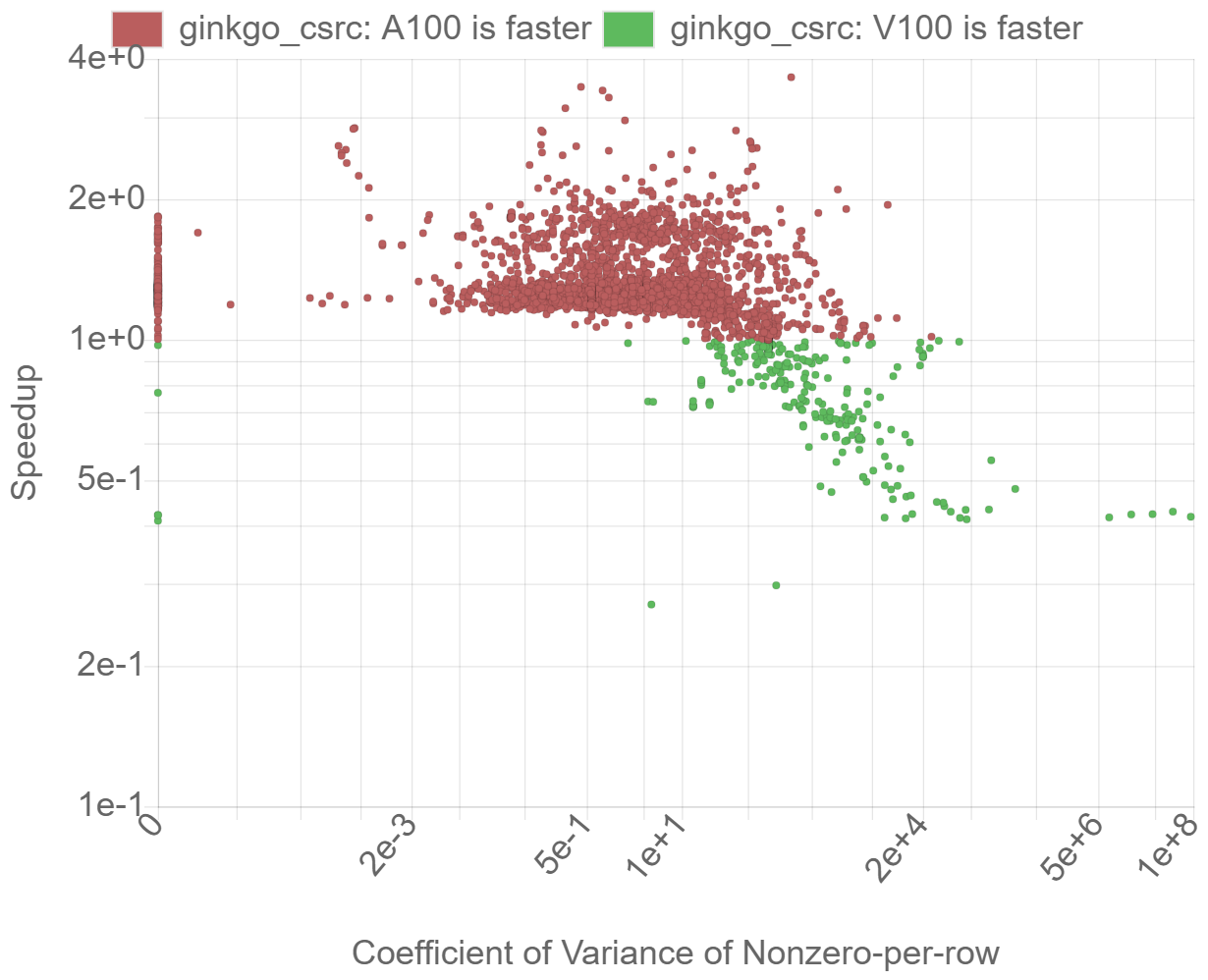}
  }
  \caption{Performance improvement of the A100 GPU over the V100 
  GPU for Ell and Classical Csr from Ginkgo with coefficient of variance of nonzero-per-row.}
  \label{fig:v100a100rcv}
\end{figure*}

\section{SpMV Performance Assessment}
\label{sec:spmvexperiments}
For the \spmv kernel performance assessment, we consider more than 2,800 matrices 
from the Suite Sparse Matrix collection~\cite{suitesparse}. Specifically, we 
consider all real matrices except for \Cref{fig:a100profile} where we require 
the matrix to contain more than 1e5 nonzero elements.  
On the A100 GPU, we run the \spmv
kernels from NVIDIA's cuSPARSE library in version 11.0 (Release Candidate) and
the Ginkgo open source library version 1.2~\cite{ginkgo}. On the older V100 GPU,
we run the same version of the Ginkgo library but use the cuSPARSE version 
10.1\footnote{cusparse\_csr and cusparse\_coo are part of CUDA version 11.0, 
CUDA 10.1 contains the conventional CSR \spmv routine}.

On the left-hand side of \Cref{fig:spmvperf}, we initially show the performance
of all considered \spmv kernels on the A100 GPU for all test matrices: each dot
represents the performance one of the \spmv kernels achieves for one test
matrix. While it is impossible to draw strong conclusions, we can observe some
CSR-based \spmv kernels exceed 200~GFLOP/s. This is consistent with the roofline
model~\cite{roofline}: if we assume every entry of a CSR matrix needs 12 bytes 
(8 bytes for the fp64 values, 4 bytes for the column index, and ignoring the 
row pointer), assume all vector entries are cached (ignoring
access to the input and output vectors), and a peak memory bandwidth of 1.4TB/s,
we end up with $2nnz\cdot \frac{1,400GB/s}{12B/nnz}\approx 230 GFLOP/s$.

On the right-hand side of \Cref{fig:spmvperf}, we show a performance
profile~\cite{10.1145/2950048} considering all \spmv kernels available in either
cuSPARSE or Ginkgo for real matrices containing more than 1e5 nonzero elements. 
As we can see, the \texttt{cusparse\_gcsr2} kernel has the largest share in terms of 
being the fastest kernel for a problem. However, \texttt{cusparse\_gcsr2} does not 
generalize well: \texttt{cusparse\_gcsr2} is more than $1.5\times$ slower than the fastest kernel for 20\%
of the problems, and more than $3\times$ slower than the fastest
kernel for 10\% of the problems. Although Ginkgo CSR \spmv kernel is not 
the fastest choice for as many problems as the \texttt{cusparse\_csr} and the 
\texttt{cusparse\_gcsr2} kernels, Ginkgo CSR generalizes better, and virtually 
never more than $2.5\times$ slower than the fastest kernels among all matrices. 
The kernels providing the best performance portability across all matrix
problems are the COO \spmv kernels from Ginkgo and cuSPARSE: only for 10\% of
the problems are they more than $1.4\times$ slower than the fastest kernel. As
expected, the SELLP \spmv kernel does not generalize well, but it is the fastest
kernel for 20\% of the problems.

We then evaluate the performance improvements when comparing the \spmv kernels
on the newer NVIDIA A100 GPU and the older NVIDIA V100 GPU. In
\Cref{fig:v100a100spmv}, we visualize the speedup factors for NVIDIA's cuSPARSE
library (left-hand side) and the Ginkgo library (right-hand side). In the first
row of \Cref{fig:v100a100spmv} we focus on the CSR performance. As expected, the
CSR \spmv achieves higher performance on the newer A100 GPU. For both libraries,
cuSPARSE and Ginkgo, the CSR kernels achieve for most matrices a 1.7$\times$
speedup on the A100 GPU -- which reflects the bandwidth increase. However, for
many matrices, the speedup exceeds 1.7$\times$. For Ginkgo, the acceleration of
up to $8\times$ might me related to larger caches on the A100 GPU allowing for 
more
efficient caching of the input vector entries. The even larger speedups of up to
$40\times$ for the cuSPARSE CSR kernel are likely only in part coming from the
newer hardware: cuSPARSE 11.0 introduces new sparse linear algebra kernels 
(generic API) that
are different in design and interface compared to cuSPARSE 9\~10.2 traditional 
kernels~\cite{cuda11}.
The analysis of the performance improvement for the COO kernels is presented in
the second row of \Cref{fig:v100a100spmv}. For matrices with less than 500,000
nonzeros, the performance improvements are about 1.3$\times$ for the Ginkgo
library and $2.1\times$ for the cuSPARSE library. For matrices with more than
500,000 nonzero elements, we observe a sudden increase of the speedup to
$>5\times$ for the cuSPARSE COO kernel, which is likely linked to a algorithm 
improvement.
In the third row of \Cref{fig:v100a100spmv}, we visualize the performance
improvements for Ginkgo's ELL and hybrid formats that do not have a direct
counterpart in cuSPARSE 11.0. Again, we see that the A100 provides for most
matrices about $1.4\times$ higher performance, however, especially for the ELL
\spmv kernel, there are several matrices where the V100 achieved higher
performance.

To investigate these singularities further, we correlate the performance 
improvement in \Cref{fig:v100a100rcv} to the coefficient of variance of the 
nonzero-per-row metric -- that is the ratio between the variance of the 
nonzero-per-row metric and the mean of the nonzero-per-row metric. Given the 
strategy Ginkgo's ELL kernel balances the work, larger coefficient of variance 
tend to introduce small data reads that perform poorly on the A100 GPU. 
Even more visible is this effect for the CSR\_C \spmv kernel, the classical 
row-parallelized CSR \spmv kernel, see the right-hand side in 
\Cref{fig:v100a100rcv}: Irregular sparsity patterns resulting in frequent loads 
of small data arrays and reflected in large coefficient of variance result in 
poor performance of the A100 GPU. Remarkably, the V100 can better deal with the 
frequent access to small data arrays. Ginkgo's CSR \spmv automatically chooses 
between the classical CSR\_C kernel providing good performance for regular 
sparsity patterns and the load-balancing CSR\_I kernel~\cite{csri} providing 
good performance for unbalanced sparsity patterns.

\section{Krylov Solver Performance Assessment}
\label{sec:krylov}
Krylov methods are among the most efficient algorithms for solving large and
sparse linear systems.
When applied to a linear system $Ax=b$ (with the sparse coefficient matrix $A$,
right-hand side $b$, and unknown $x$)
Krylov solvers started with an initial guess $x_0$ produce a sequence of vectors
$x_1, x_2, x_3,\ldots$ that, in general, progressively reduce
the norm of the residuals $r_k=b-Ax_k$, eventually yielding an acceptable
approximation to the solution of the system~\cite{Anzt2017a}.

Algorithmically, every iteration of a Krylov solver is composed of a (sparse)
matrix vector product to generate the new search direction, an
orthogonalization procedure, and the update of the approximate solution and the
residual vector. In practice, Krylov methods are often enhanced with
preconditioners to improve robustness and convergence~\cite{Anzt2017a}.
Ignoring the preconditioner, every iteration can be composed of level-1 BLAS
routines (vector operations) and a level-2 BLAS in the form of a sparse matrix
vector product (\spmv). All these components -- and in virtually all cases also
the preconditioner application -- are memory bound operations.

When upgrading from the V100 to the A100 GPU, for the vector updates and
reduction operations, we can expect performance
improvements corresponding to the bandwidth improvements observed in
\Cref{sec:bandwidth}. For the basis-generating \spmv kernel, the improvements
are problem-dependent and may even exceed the 1.7$\times$ bandwidth
improvement, see \Cref{sec:spmvexperiments}. The total solver speedup then
depends on how the Krylov method composes the \spmv and vector operations, and
how these components contribute to the overall runtime.

\begin{figure}
    \centering
    \includegraphics[width=.45\textwidth]{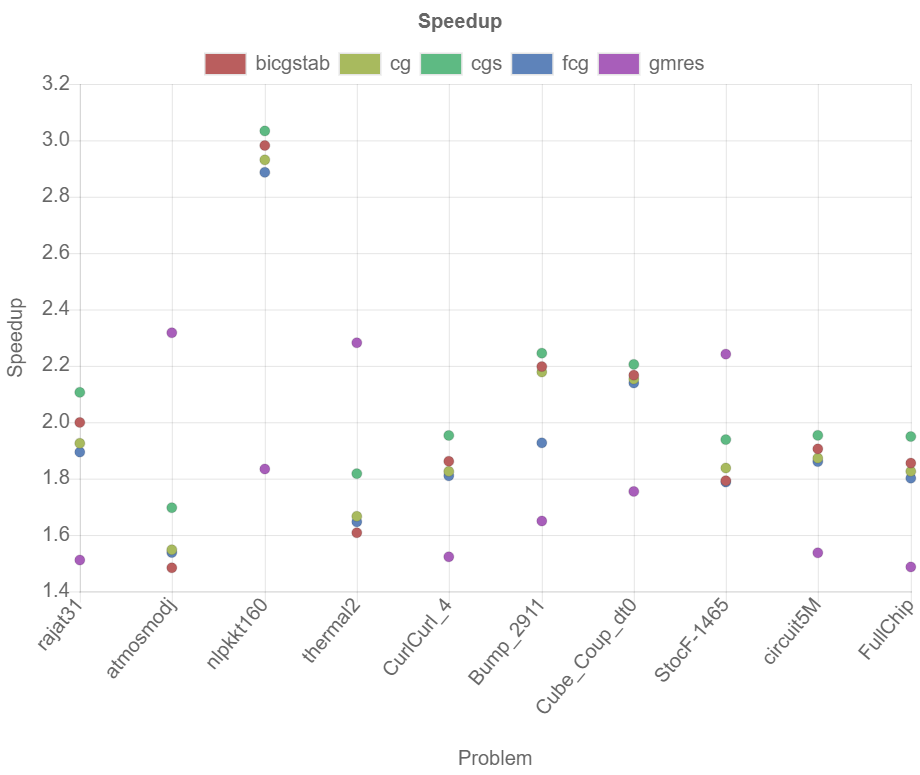}
    \caption{Krylov solver acceleration when upgrading from the V100 GPU to the
    A100 GPU.}
    \label{fig:solverspeedup}
\end{figure}

In \Cref{fig:solverspeedup}, we visualize the performance improvement observed
when upgrading from the V100 to the A100 GPU. We select 10 test matrices that
are large in terms of size and nonzero elements, different in their
characteristics and representative for different real-world applications. The
Krylov solvers are all taken from the Ginkgo library~\cite{ginkgo}, the \spmv
kernel we employ inside the solvers is Ginkgo's COO \spmv kernel. For most test
problems, we actually observe larger performance improvements than what the
bandwidth ratios suggest. We also observe that if focusing exclusively on a
single test problem, the speedup factors for the Krylov methods based on short 
recurrences (i.e. bicg, cg, cgs, fcg) are all almost identical. These methods 
are all very similar in
design, and the \spmv kernel takes a similar runtime fraction. The gmres
algorithm is not based on short recurrences but builds up a complete search
space, and every new search direction has to be orthogonalized against all
previous search directions. As this increases the cost of the orthogonalization
step -- realized via a classical Gram-Schmidt algorithm -- the \spmv accounts 
for a smaller fraction of the algorithm runtime. As a
result, the speedup for gmres is often different than the speedup for the other
methods. However, also for gmres the speedup can exceed the 1.7$\times$
bandwidth improvement as the A100 features larger caches that can be a
significant advantage for the orthogonalization kernel. Overall, we observe that
for most scenarios we tested, the Krylov solver executes on the A100 GPU more
than 1.8$\times$ faster than on the V100 GPU.

\section{Conclusion}
\label{sec:conclusion}
In this paper, we assessed the performance NVIDIA's new A100 GPU achieves for 
sparse linear algebra operations. As most of these algorithms are memory bound, 
we initially present results for the STREAM bandwidth benchmark, then provide a 
very detailed assessment of the sparse matrix vector product performance for 
both NVIDIA's cuSPARSE library and the Ginkgo open source library, and 
ultimately run complete Krylov solvers combining vector operations with sparse 
matrix vector products and orthogonalization routines. Compared to the 
predecessor, the A100 GPU provides a $1.7\times$ higher memory 
bandwidth, achieving almost 1.4 TB/s for large input sizes. The larger caches 
on the A100 allow for even higher performance improvements in complex 
applications like the sparse matrix vector product. Ginkgo's Krylov iterative 
solver run in most cases more than 1.8$\times$ faster on the A100 GPU than on 
the V100 GPU.

\section*{Acknowledgments}
This work was supported by the ``Impuls und Vernetzungsfond'' of the Helmholtz 
Association under grant VH-NG-1241 and by the Exascale Computing Project 
(17-SC-20-SC), a
collaborative effort of the U.S. Department of Energy Office of Science and the
National Nuclear Security Administration.
The authors would like to thank the Steinbuch Centre for Computing (SCC) of the 
Karlsruhe Institute of Technology for providing access to an NVIDIA A100 GPU.

\bibliographystyle{plain} 
\bibliography{references} 
\onecolumn
\section*{Appendix: NVIDIA V100/A100 Tensor Core GPU Architecture}
\label{app:a100}

\begin{table}[!h]
\centering
\begin{tabular}{||l|l|l||} 
    \hline
    Features & V100 & A100 \\
    \hline\hline
    GPU Architecture & NVIDIA Volta & NVIDIA A100 \\
    SMs & 80 & 108 \\
    TPcs & 40 & 54\\
    FP32 Cores/SM & 64 & 64 \\
    FP64 Cores/SM & 32 & 32 \\
    INT32 Cores/SM & 64 & 64 \\
    GPU Boost Clock & 1530 MHz & 1410 MHz \\
    Peak FP16 TFLOPS & 31.4 & 78 \\
    Peak FP32 TFLOPS & 15.7 & 19.5 \\
    Peak FP64 TFLOPS & 7.8 & 9.7 \\
    Texture Units & 320 & 432 \\
    Memory Interface & 4096-bit HBM2 & 5120-bit HBM2 \\
    Memory Data Rate & 877.5 MHz DDR & 1215 MHz DDR \\
    Memory Bandwidth & 900 GB/sec & 1555 GB/sec \\
    L2 Cache & 6144 KB & 40960 KB \\
    Shared Memory Size/SM & 96 KB & 164 KB \\
    Register File Size & 256 KB & 256 KB \\
    Transistors & 21.1 billion & 54.2 billion \\
    \hline
\end{tabular}
\label{tab:a100specs}
\caption{Technical characteristics of NVIDIA's V100 and A100 GPU architecture in comparison to its predecessors~\cite{a100}.}
\end{table}


\end{document}